\documentclass[twocolumns]{emulateapj}
\usepackage{natbib}
\usepackage{amssymb}
\usepackage{amsmath}
\def\be{\begin{eqnarray}}
\def\ee{\end{eqnarray}}
\usepackage{comment}
\usepackage{color}
\usepackage{graphicx,subfigure}
\usepackage{enumerate}
\usepackage{rotating}
\usepackage{longtable}

\shorttitle{Radio follow-up of GW170608}

\shortauthors{K.~Artkop, R.~Smith, A.~Corsi et al.}

\begin{document}

\title{Radio follow-up of a candidate $\gamma$-ray transient in the sky localization area of GW170608}
\author{Kyle Artkop\altaffilmark{1}, Rachel~Smith\altaffilmark{1}, Alessandra~Corsi\altaffilmark{1}, Simona~Giacintucci\altaffilmark{2}, Wendy~M. Peters\altaffilmark{2}, Rosalba~Perna\altaffilmark{3,4}, S.~Bradley~Cenko\altaffilmark{5,6}, Tracy~E.~Clarke\altaffilmark{2}}
\altaffiltext{1}{Department of Physics and Astronomy, Texas Tech University, Box 1051, Lubbock, TX 79409-1051, USA; e-mail: alessandra.corsi@ttu.edu}
\altaffiltext{2}{Remote Sensing Division, Naval Research Laboratory
Washington, DC 20375-5320, USA}
\altaffiltext{3}{Department of Physics and Astronomy, Stony Brook University, Stony Brook, NY 11790, USA}
\altaffiltext{4}{Center for Computational Astrophysics, Flatiron Institute, New York, NY 10010, USA}
\altaffiltext{5}{Astrophysics Science Division, NASA Goddard Space Flight Center, Mail Code 661, Greenbelt, MD 20771, USA}
\altaffiltext{6}{Joint Space-Science Institute, University of Maryland, College Park, MD 20742, USA}

\begin{abstract}
\label{abstract}
After the identification of a candidate $\gamma$-ray transient in the error region of the binary black hole (BBH) merger GW150914 by the \textit{Fermi} satellite, the question of whether BBH mergers can be associated to electromagnetic counterparts remains highly debated. Here, we present radio follow-up observations of GW170608, a BBH merger that occurred during the second observing run (O2) of the Advanced Laser Interferometer Gravitational-wave Observatory (LIGO). Our radio follow up focused on a specific field contained in the GW170608 sky localization area, where a candidate high-energy transient was detected by the \textit{Fermi} Large Area Telescope (LAT). We make use of data collected at 1.4\,GHz with the Karl G. Jansky Very Large Array (VLA), as well as with the VLA Low-band Ionosphere and Transient Experiment (VLITE). Our analysis is sensitive to potential radio afterglows with luminosity densities $L_{\rm 1.4\,GHz}\gtrsim 6\times10^{28}$\,erg\,s$^{-1}$\,Hz$^{-1}$. In the most optimistic theoretical models, $\approx 20\%$ of BBH events occurring in massive hosts could be associated with outflows as radio luminous as this.  Although we find no evidence for the presence of a radio counterpart associated with the LAT candidate in the GW170608 error region, our analysis demonstrates the feasibility of future radio follow-up observations of well localized BBHs. Comparing our radio upper-limits with theoretical expectations for the radio afterglows  potentially  associated  with  jets  launched  in
BBH mergers, we find that for jets  of  energy $\approx 10^{49}$\,erg seen on-axis, only  jet  angles $\theta_{jet}\gtrsim 40^{\circ}$ are compatible with the observations.
\end{abstract}

\keywords{\small gravitational waves --- radiation mechanisms: general  --- radio continuum: general}

\section{Introduction}
\label{intro}
The recent direct detection of gravitational waves (GWs) from stellar-mass binary black holes (BBHs) by Advanced LIGO (Laser Interferometer Gravitational Wave Observatory) and Virgo has opened a new era in the study of the most exotic objects in the stellar graveyard \citep{Abbott2016a,Abbott2016b,Abbott2017a,Abbott2017b,Abbott2017c,AbbottCatalog}. In 2017, the direct detection of GWs from BBHs was awarded the Nobel Prize in Physics. Following that, the remarkable discovery of GW170817 \citep{Abbott2017d}, the first binary neutron star (NS) merger detected in GWs with an electromagnetic (EM) counterpart \citep{Abbott2017e}, helped demonstrate the wide impact of multi-messenger astronomy on a variety of fields, including nucleosynthesis, extreme states of nuclear matter, and cosmology \citep[e.g.,][]{Abbott2017f,Abbott2018}.

According to traditional paradigms, one would not expect to detect EM emission accompanying BBH mergers as the immediate circmum-merger environment is expected to be rather clean post merger (i.e., lacking ejecta mass in the merger site). 
However, the intriguing \textit{Fermi}/GBM $2.9\sigma$-significance detection of a short burst of $\gamma$-rays (GRB) possibly associated with the BBH GW150914 \citep{Connaughton2016,Connaughton2018}, has spurred the investigation of several new theoretical scenarios for EM emission from BBHs \citep[e.g.,][]{Liebling2016,Loeb2016,Perna2016,Tagawa2016,Bartos2017,Dai2017,deMink2017,Dolgov2017,Fedrow2017,Ioka2017,Kelly2017,Kimura2017,Shapiro2017,Dorazio2018,Khan2018}. As of today, the question of whether BBH mergers can be accompanied by a release of
energy in the form of a relativistic outflow powering a $\gamma$-ray counterpart remains open and highly debated \citep[e.g.,][]{Bhalerao2017,Stalder2017,Verrecchia2017,Perna2018,Perna2019}.  Certainly, the detection of an EM counterpart to BBHs would start a revolution in the way we have traditionally thought of stellar evolution and accretion processes in astrophysics. Observationally, if at least some BBHs are associated with GRBs, one can hope to strengthen the significance of a potential association by searching  for their broad-band afterglows \citep[e.g.,][]{Kasliwal2016,Morsony2016,Murase2016,Palliyaguru2016,Veres2016,Yamazaki2016,Perna2018,Perna2019}. Indeed, the detection of an afterglow from a candidate GRB counterpart to the GW signal would provide a much improved localization (compared to the GW and $\gamma$-ray localizations), and constrain key physical parameters such as total kinetic energy of the ejecta, density of the surrounding medium, opening angle of the ejecta, and viewing angle. 

Here, we report on a search for a potential radio afterglow from the BBH merger observed by the two LIGO detectors on 8 June 2017 UT, henceforth referred to as GW170608 \citep{Abbott2017c}. This is the least massive BBH merger detected by LIGO to date, with component masses of $12_{-2}^{+7} M_{\odot}$ and $7_{-2}^{+2} M_{\odot}$ \citep[90\% credible intervals;][]{Abbott2017c}, and a luminosity distance of $340_{-140}^{+140}$\,Mpc \citep{Abbott2017c}.  

Several EM observatories around the globe have carried out follow-up observations of at least a portion of the sky localization area of GW170608. Our radio observations with the Karl G. Jansky Very Large Array (VLA), executed under program VLA/16A-206 (PI: Corsi), have targeted the location of a potential high-energy $\gamma$-ray transient detected by the \textit{Fermi}/LAT at the 3.5$\sigma$ significance level, $\approx 1200$\,s after the GW trigger \citep{Omodei2017}. 
We supplemented our GHz VLA observations with lower frequency radio observations from the VLA Low-band Ionosphere and Transient Experiment (VLITE). 

Here, we present our analysis of the collected VLA and VLITE data, and discuss their implications for models of relativistic ejecta from BBH mergers. 
Our paper is organized as follows: In Section \ref{sec:obssumm}, we summarize the observations that led to the identification of a candidate $\gamma$-ray transient in the error area of GW170608. In Section \ref{sec:obs}, we describe our radio follow-up observations and data reduction. In Section \ref{sec:results}, we show how no convincing radio transient was found in association with the \textit{Fermi}/LAT candidate event, and discuss the implications of these results. Finally, in Section \ref{sec:conclusion}, we summarize and conclude. 

\begin{table}
\begin{center}\begin{footnotesize}
\caption{X-ray sources identified by \textit{Swift}/XRT in the error region of GW170608. Columns are, from left to right: Source number, R.A., Dec., localization error, and 0.3-10\,keV flux. See text for discussion.\label{tb:SwiftSources}}
\begin{tabular}{lcccc}
\hline\hline
  &  R.A.\,Dec.         & Pos. Err. & $F_X$ \\ 
                  & (hh:mm:ss\,deg:mm:ss)   &  \arcsec            & (erg\,s$^{-1}$\,cm$^{-2}$) \\
\hline
XRT-s1    &08:33:03.19\,+43:12:46.8& 8.6 & 6.1$\times 10^{-13}$\\
XRT-s2    &08:32:07.70\,+43:29:51.0& 5.7 & 1.7$\times 10^{-13}$\\
XRT-s3    &08:31:31.70\,+43:25:48.0& 6.3 & 5.4$\times 10^{-13}$\\
XRT-s4    &08:32:52.18\,+43:23:08.9& 6.0 & 3.7$\times 10^{-13}$\\
XRT-s5    &08:34:07.27\,+43:19:09.1& 7.2 & 4.0$\times 10^{-13}$\\
XRT-s6   &08:33:42.79\,+43:17:21.1& 5.0 & 2.2$\times 10^{-13}$\\
XRT-s7    &08:31:12.53\,+43:25:43.7& 7.6 & 5.1$\times 10^{-13}$\\
\hline
\end{tabular}
\end{footnotesize}\end{center}\end{table}

\section{GW170608 detection and candidate $\gamma$-ray transient identification}
\label{sec:obssumm}
The merger associated with the BBH GW170608 was detected by LIGO as a two-detector coincident event at $T_0=$02:01:16.49 UT on 2017 June 8. An alert was issued to EM observing partners $\approx 13.5$\,hrs later, with a sky localization area spanning $\approx 860$\,deg$^2$
\citep[90\% credible region; ][]{Abbott2017c,Abbott2017g}. In response to this alert, several observatories searched their data for potential transients located within the large GW sky area\footnote{GRB Coordinates
Network Circulars (GCNs) related to the EM follow-up of GW170608 are archived at
https:$//$gcn.gsfc.nasa.gov$/$other$/$G288732.gcn3
.}. Among these, the \textit{Fermi}/LAT team performed a search for high-energy transients in data collected during a 10\,ks time interval after $T_0$ \citep{Omodei2017}. The highest significance ($\approx 3.5\sigma$ level) excess found was found about 1200\,s after $T_0$ at R.A.=08\,h\,32\,m\,26.400\,s  and Dec=+43\,d\,23\,m\,24.00\,s (J2000), with a localization uncertainty of 0.24\,deg (at 90\% confidence). The location of this candidate $\gamma$-ray counterpart was occulted by the Earth before $T_0+1200$\,s.  No \textit{Fermi}/GBM counterpart was found for the candidate \textit{Fermi}/LAT transient. The \textit{Fermi}/GBM 3-$\sigma$, 1-second-averaged flux upper limit was in the range $(4.7-5.6)\times10^{-7}$\,erg\,s$^{-1}$\,cm$^{-2}$ between about $T_0+85$\,s and $T_0+1$\,hr \citep{Goldstein2017}, which at a distance of 340\,Mpc corresponds to an isotropic $\gamma$-ray energy flux upper-limit of $(6.5-7.7)\times10^{48}$\,erg\,s$^{-1}$.

The \textit{Swift}/XRT team performed a four-pointing follow up of the possible LAT source, with observations spanning a time interval between $\approx 0.7$\,d\ and $\approx 1$\,d after the GW trigger \citep{Evans2017}. Seven X-ray sources were identified in the observed fields,  two of which were previously catalogued (referred to as XRT sources 4 and 5; see Table \ref{tb:SwiftSources}). None of these sources showed evidence for flux variations over \textit{Swift}'s observation period, and none exhibited signs of outburst
compared to previous observations. Thus, all of these seven sources were deemed unlikely to being associated with GW170608. For completeness and comparison with our VLA/VLITE results, we report in Table \ref{tb:SwiftSources} the R.A., Dec. (J2000), localization error, and 0.3-10\,keV fluxes of these \textit{Swift}/XRT sources.
\textit{Swift}/UVOT observations further revealed that \textit{Swift}/XRT sources 3, 6, and 7 were already present in the SDSS catalog, and none of them exhibited signs of fading over a time interval of $\approx 0.7-1$\,d after the GW trigger \citep{Emery2017}. 

Following the identification of the \textit{Fermi}/LAT high-energy candidate transient, several other optical telescopes observed its LAT localization area. The Nanshan One-meter Wide field Telescope (NOWT) team observed the field of the LAT transient in the $R$-band around 15:30:19 UT on 09 June 2017 \citep{Xu2017}. They noted the existence of optical counterparts to \textit{Swift}/XRT sources 1, 3, 6 and 7 in PanSTARRS, DSS-II, and/or SDSS, and excluded evidence for significant optical flux variability of these sources. No optical counterpart was found for source 2 down to $R>19.2$\,mag. 

The Arizona Transient Exploration and Characterization (AZTEC) team observed the field of the \textit{Fermi}/LAT candidate in $i$ band with the Large Binocular Camera (LBC) mounted on the
Large Binocular Telescope (LBT) on beginning on 2017
June 10 \citep[about 2 days post GW trigger;][]{Fong2017}. The observations covered $\approx 88\%$ of the \textit{Fermi}/LAT localization region, and included the locations of \textit{Swift}/XRT sources 1-4 (see Table \ref{tb:SwiftSources}). No new sources were found in or around the positions of XRT sources 1-4, and no evidence for variability was reported. This same team observed the  field of the \textit{Fermi}/LAT candidate also with the Wide Field Camera (WFCAM) on the 3.8-m United
Kingdom Infrared Telescope \citep[UKIRT;][]{Fong2017a}, beginning on 2017 Jun 9.2 UT ($\approx 1$\,d after the GW trigger) and  2017 June 10.2 ($\approx 2$\,d after the GW trigger). The first epoch covered the full 90\% \textit{Fermi}/LAT localization region, and the positions of the seven
\textit{Swift}/XRT sources. The second epoch covered 75\% of the \textit{Fermi}/LAT
localization area, and the positions of XRT sources 2-7 (see Table \ref{tb:SwiftSources}). Again no new sources or evidence for variability were found \citep{Fong2017a}.

The LIGO error region of GW170608 entered the field of view of the HAWC Cherenkov array about 1\,d after $T_0$. No evidence for transients at a significance level of $\gtrsim 3\sigma$ was reported in the 0.5-100\,TeV band \citep{Smith2017a}. 

The J-GEM collaboration covered almost the entire error region of the \textit{Fermi}/LAT transient candidate, and the location of all \textit{Swift}/XRT sources 1-7, with the 1.05-m Kiso Schmidt telescope. Observations were performed in $i$-band on 2017 June 9.483, 9.489, 9.527, and 9.529\,UT, approximately 1.4\,d 
after the GW detection, at a median $5\sigma$ depth of $i=15.6$\,mag (AB system). No new transients were found and lack of variability was established for \textit{Swift}/XRT source 7 \citep{Morokuma2017}.

Finally, the Pan-STAARS1 telescope observed the \textit{Fermi}/LAT error circle for 11 nights beginning a day after the GW detection. No new sources were detected within the 90\% \textit{Fermi}/LAT localization area of the candidate $\gamma$-ray transient at a depth of $i,z \approx 18.5$\,mag (within 1\,d post merger) and $i,z \approx 20.5$   \citep[daily stacked limits up to 5\,d after $T_0$, see][]{Smith2017b}. 

\section{Radio follow-up observations}
\label{sec:obs}
\subsection{VLA observations and data reduction}
We observed the field of the  \textit{Fermi}/LAT candidate $\gamma$-ray counterpart to GW170608 with the VLA at a central frequency of 1.4\,GHz, and with a nominal bandwidth of 2\,GHz \citep[][]{Corsi2017}.  Three observations were carried out on 2017 June 12, June 29, and August 18\,UT, all with the VLA in its C configuration. We imaged an area with a nominal $\approx 0.54$\,deg FWHM primary beam centered around the position of the LAT excess (R.A.=08\,h\,32\,m\,30\,s, Dec=+43\,d\,24\,m\,00\,s), fully enclosing the 90\% confidence error area of the LAT localization. 

The VLA data were calibrated using the VLA automated pipeline in CASA \citep{McMullin2007}. After calibration, we inspected the data for RFI and applied any necessary flagging. Specifically, data were heavily affected by RFI (which is not uncommon at L-band), reducing the effective bandwidth to $\approx 40\%$ of the nominal $\approx 1$\,GHz bandwidth of the L-band receiver. Images of the field were formed using the CLEAN algorithm in interactive mode. The FWHM of the major axis of our synthesized beam ranged between $\approx 14$\,\arcsec--$17$\,\arcsec, consistent with expectations for the VLA in its C configuration. While forming images, we included primary beam corrections to account for the shape of the primary beam up to a region extending to 20\% of the power radius ,which is the standard option in CLEAN, translating to images of angular diameter $\approx 0.7$\,deg.  Over the three epochs, we reached a typical central image RMS of $\approx 45$\,{$\mu$Jy}. The last was estimated with IMSTAT using a circle of radius 60\arcsec\, from the center of the images. Our image rms is $\approx 2\times$ higher than expected based on the time spent on-source ($\approx 1$\,h and 20\,min), and the actual bandwidth reduction due to RFI. This is due to limited dynamic range of our images related to the presence of bright sources in the crowded field.

After calibration, flagging, and imaging, we visually inspected the images and identified sources with signal-to-noise ratio ${\rm SNR}\,\gtrsim 10$, or flux densities greater than $\approx 450\,\mu$Jy at 1.4\,GHz. At the distance of GW170608 ($\approx 340$\,Mpc), this flux density limit corresponds to a radio luminosity density of $\approx 6\times10^{28}$\,erg\,s$^{-1}$\,Hz$^{-1}$. 
Our results are reported in Table \ref{tb:VLASources}. Source coordinates were calculated with the IMFIT algorithm, using a circular region of radius 10\arcsec (i.e., of diameter comparable to the FWHM of the synthesized beam), centered around the source position determined through visual inspection. We then used the IMSTAT algorithm to determine the peak flux density of each source within a circular region of radius 10\arcsec\,centered around the coordinates obtained via IMFIT. Peak flux errors were calculated by adding in quadrature the RMS noise corrected for primary beam effects (by rescaling the central image RMS by the primary beam correction at the location of each source), and a nominal 5\% absolute flux calibration error. Finally, position errors were calculated by dividing the FWHM of the semi-major axis of the synthesized beam by the source signal-to-noise ratio (SNR; peak flux density divided by peak flux density error).  

\subsection{VLITE observations and data reduction}
The VLITE \citep{Clarke2016} is a commensal, low-frequency system on the VLA that runs in parallel with nearly all observations above 1\,GHz. VLITE provides real-time correlation of the signal from a subset of VLA antennas using the low band receiver system \citep{Clarke2011} and a dedicated DiFX-based software correlator \citep{Deller2007}. The VLITE system processes 64 MHz of bandwidth centered on 352 MHz, but due to strong radio frequency interference (RFI) in the upper portion of the band, the usable frequency range is limited to an RFI-free band of $\sim 40$ MHz, centered on 338 MHz. 

VLITE was operational with 15 working antennas during the VLA 1.4\,GHz observations of the \textit{Fermi}/LAT candidate transient on UT 2017 Aug 18. The VLITE data collected during this epoch were processed using a dedicated calibration pipeline, which is based on a combination of the Obit \citep{Cotton2008} and AIPS \citep{VanMoorsel1996} data reduction packages. The calibration pipeline uses standard automated tasks for the removal of RFI and follows common techniques of radio-interferometric data reduction, including delay, gain and bandpass calibration  \citep[for details on the pipeline data reduction see][]{Polisensky2016}. The flux density scale is set using Perley \& Butler (2017) and residual amplitude errors are estimated to be less than $20\%$ (Clarke et al., in preparation). The data were imaged using wide-frequency imaging algorithms in Obit (task MFimage), by covering the the full primary beam with facets and placing outlier facets on bright sources out to a radius of $20^{\circ}$. Small clean masks are placed on the sources during the imaging process to reduce the effects of CLEAN bias. The pipeline runs two imaging and phase self-calibration cycles before a final image is created. 

To improve the quality of the image, the pipeline-calibrated data were re-imaged by 
hand with MFImage using a higher number of cleaning iterations, which reduced the RMS by a factor 
of approximately two. The final VLITE image has a restoring beam of $73^{\arcsec}\times43^{\arcsec}$ and a RMS noise of $\approx 2.5$ mJy\,beam$^{-1}$ (1$\sigma$). The image was used to search  for counterparts  to  our  VLA
sources.   These are reported in Table \ref{tb:VLITESources} with their position, flux, and errors. Flux densities were measured in AIPS using a Gaussian fit (JMFIT) and corrected for the primary beam attenuation using factors appropriate for VLITE (Polisensky et al. in preparation).  Flux density errors include local image noise and flux scale uncertainty.

\section{Results and Discussion}
\label{sec:results}
\subsection{Observational results}
\begin{table*}
\begin{center}\begin{footnotesize}
\caption{VLITE counterparts to VLA sources. Columns are, from left to right: Source number, VLA coincident source number, R.A., Dec., position error, and flux density as measured in the VLITE image. S4/S6 and S10/S11 are detected as one single source by VLITE.\label{tb:VLITESources}}
\begin{tabular}{ccccc}
\hline\hline
  &VLA\,\#&  R.A.\,Dec.         & Pos. Err. & $F_{\rm 338\,MHz}$ \\ 
  && (hh:mm:ss\,deg:mm:ss)&  (\arcsec) & (mJy) \\
\hline
V1  &S1&08:33:16.53\,+43:23:50.1& 9.43 & $15.1 \pm\,3.9$\\
V2  &S2&08:31:25.95\,+43:25:13.4& 7.30 & $523.9 \pm\,104.8$\\
V3  &S3&08:32:43.15\,+43:14:38.1& 7.37 & $100.1 \pm\,20.2$\\
V4  &S4\,+\,S6&08:32:28.64\,+43:02:56.1& 7.30 & $380.8 \pm\,76.2$\\
V5  &S5&08:31:11.47\,+43:16:08.7& 7.85 & $32.1 \pm\,6.9$\\
V6  &S7&08:31:00.19\,+43:27:27.9& 8.07 & $26.7 \pm\,5.9$\\
V7  &S10\,+\,S11&08:33:49.38\,+43:20:25.0& 7.35 & $127.2 \pm\,25.6$\\
\hline
\end{tabular}
\end{footnotesize}\end{center}\end{table*}

In our analysis of the 1.4\,GHz VLA images of the field of the \textit{Fermi}/LAT candidate $\gamma$-ray transient, we have identified a total of 25 radio sources with $SNR\gtrsim 10$, or $F_{1.4\,GHz}\gtrsim 450\,\mu$Jy ($L_{\rm 1.4\,GHz}\approx 6\times10^{28}$\,erg\,s$^{-1}$\,Hz$^{-1}$ at the distance of GW170618). These sources are listed in Table 3.  In the most optimistic scenarios for afterglows associated with BBH mergers,$\approx 20\%$ of BBH events occurring in massive (spiral or elliptical) hosts could be associated with GRB-like outflows of radio luminosity density larger than or equal to this value. \citep[see e.g. Figure 8 in ][]{Perna2018}.

A comparison of the VLA sources (Table 3) with those found by the \textit{Swift}/XRT team in the GW error region (Table \ref{tb:SwiftSources}) shows that only two of the \textit{Swift}/XRT sources are within a distance of 2\arcmin\, from any of our VLA sources. Specifically, \textit{Swift}/XRT-s1 and \textit{Swift}/XRT-s2 are located within $\approx 0.9$\arcmin\ of our VLA source S13 and S20, respectively. We note that at the distance of GW170608 ($\approx 340$\,Mpc), an angular radius of $2\arcmin$ corresponds to a physical distance of about 170\,kpc. As shown by \citet{Perna2018}, this radial distance is likely to enclose the host galaxy of $\gtrsim 70-90\%$ of BBH mergers occurring in massive hosts (see their Figure 3). Thus, searching for sources located within 2\,\arcmin of the position of our VLA sources leaves room to find not only coincident counterparts but also potential host galaxies. 

We also searched for any counterpart to our VLA sources in VLITE images. Only a few of the VLA sources were found with $SNR\gtrsim 5$ (or flux density $\gtrsim 12.5$\,mJy) in the VLITE image taken on 2017 August 18 UT at 338\,MHz. These V1-7 sources are reported in Table \ref{tb:VLITESources}. They were unresolved at the angular resolution of the image (73\arcsec\,x 43\arcsec\,).

To gain additional information on the possible nature of the 25 sources identified in our VLA images, we searched for previously known sources co-located with them. To this end, we queried the catalogs by the National VLA Sky Survey (NVSS), the VLA FIRST, and the NASA/IPAC Extra-galactic Database (NED). 
 Table 3 reports information about the closest (in terms of sky position, considering position errors) and most similar (in terms of radio flux) previously known radio source found within $2\arcmin$ of each of our VLA sources. When a previously known radio source is indeed found, we classify it as RadioS in Table 3, and report its
R.A. and Dec., radio peak flux density, and position 
uncertainty. The position uncertainty corresponds to the semi-major position uncertainty as reported in NED for all sources with an NVSS counterpart \citep{Condon1998}. For FIRST sources, we divide the FWHM
of  the  semi-major  axis  of  the  synthesized  beam  by  the
source signal-to-noise ratio, where the last takes into account the clean bias of 0.25\,mJy which is added to all peak fluxes reported in FIRST\footnote{\bf See http://sundog.stsci.edu/first/catalogs/readme.html for information about the FIRST survey.}. Because of systematic uncertainties, any position error smaller than 0.1\,\arcsec in Table 3 is set equal to this systematic uncertainty value.
If a radio source is not found within 2\,\arcmin, we report in the Table 3 only information about the classification of the closest known source found in NED (see sources S17-S20 and S24-25 in Table 3). 

Overall, all of the 25 radio sources we found in our VLA images are associated with
previously known sources located within 2\,\arcmin. All but six of the VLA sources have an NVSS or FIRST counterpart within 2\,\arcmin. The known sources found in association with our VLA sources S4, S19, and S24-S25 show offsets larger that the estimated position errors, so they could be either unrelated sources (as most likely for S25), or potential host galaxies.  Incidentally we note that the radio source coincident with S22 is a galaxy located at a distance much father than that of GW170608.
No NVSS/FIRST sources are found in coincidence with our VLA S17-S20 and S24-25, and this is not surprising considering that the VLA 1.4\,GHz peak flux densities of these sources are close or below the FIRST/NVSS catalog completeness limits of $\approx 0.75$\,mJy / $2.5$\,mJy \citep{White1997,Condon1998}.

None of the 25 sources identified in our VLA images showed any evidence for significant variability within the timescales of our 3 VLA epochs, which covered days 4, 21, and 67 since the BBH merger. Although large uncertainties affect predictions for radio emission from BBHs, the lack of variability on these timescales strongly suggests that all of the identified VLA sources are unrelated to possible afterglow-like emission from the \textit{Fermi}/LAT candidate. Specifically, in the standard synchrotron model for radio emission from fast GRB ejecta, one would expect a post-peak temporal evolution of the optically thin  radio afterglow such that $F_{\rm 1.4\,GHz}\propto (t/t_{\rm peak})^{-0.75}-(t/t_{\rm peak})^{-1.1}$ \citep[][and references therein]{Perna2018}, which would ensure flux variations of a factor of $\gtrsim 2$ between successive epochs of our follow-up campaign. 

\subsection{Constraints on the presence of a relativistic jet}
Given the lack of a radio counterpart to the candidate \textit{Fermi}/LAT transient, our VLA observations constrain the flux density of any potential radio afterglow to $F_{1.4\,GHz}\lesssim 450\,\mu$Jy. We thus explore whether this constraint can rule out at least some portions of the parameter space allowed for a relativistic jet potentially produced in association with the GW170608 merger. 
Light curves for jets propagating in the clean environment expected around BBH mergers
(i.e. one that lacks ejecta mass in the merger site) have been computed by \citet{Perna2019}. We use their online library\footnote{http://www.astro.sunysb.edu/rosalba/EMmod/models.html}
to derive the model radio luminosities at 1.4\,GHz, and compare them with our limits. The models are
characterized by the jet energy $E_{\rm jet}$ and the angle $\theta_{\rm jet}$ over which the bulk of the
energy is distributed.  For any of these parameters, the luminosity is then a strong function of the viewing angle
with the jet axis, and the time of the observation. We set the last to 4\,d after the merger for comparison with our VLA observations, as this is the most constraining epoch.  For the viewing angle, we assume that our line of sight is along the jet axis  (given the potential detection of a \textit{Fermi}/LAT high-energy transient and the GW selection effects which favor a face-on orientation for detection). This ``down-the-barrel'' assumption for the line-of-sight is the	most constraining one. In fact, viewing angles with $\theta_{\rm obs} > \theta_{\rm
jet}$ would lead to weaker constraints on the energetics than the ones
we present in what follows (in other words, higher jet energies would be allowed for larger viewing angles).

\begin{figure}
\begin{center}
\includegraphics[width=8cm]{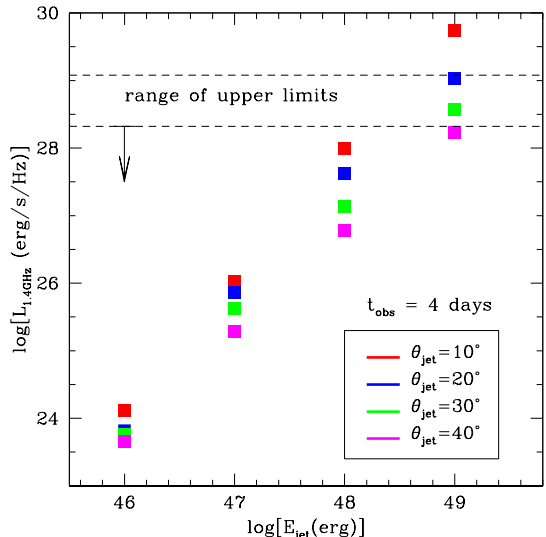}
\vspace{-0.8in}
\caption{1.4 GHz luminosity density at 4\,d after the merger, from a relativistic
  jet with Lorentz factor $\Gamma=100$ seen on axis. Models are considered for
a range of jet energies $E_{\rm jet}$ and opening angles $\theta_{\rm jet}$ of the jet. The dashed lines
mark the region of upper limits on the 1.4~GHz luminosity density derived from our VLA observations given the uncertainties on the estimated  distance of GW170608. See text for discussion. \label{fig:Lradio}}
\end{center}
\end{figure}

Figure~\ref{fig:Lradio} shows a comparison of the on-axis jet luminosity for a range of relativistic jets 
(Lorentz factor of $\Gamma=100$ \footnote{For the same model parameters, lower Lorentz factors yield lower luminosities. E.g, for $\Gamma=10$ the brightest point in Fig.~\ref{fig:Lradio} would be 80\% of the one for $\Gamma=100$. })
with energies
varying between $10^{46}-10^{49}$~ergs, and opening angles between $10^\circ-40^\circ$. These models
were chosen to bracket the energetics of the candidate {\em Fermi} counterpart to GW150914 \citep{Connaughton2016,Connaughton2018} which had an isotropic inferred energy of $\approx 10^{49}$~ergs, which translates into a jet energy
of $10^{49}(1-\cos\theta_{\rm jet})$\,ergs.  The dashed lines in  Fig.~\ref{fig:Lradio}  mark the range of observational upper limits
on the radio luminosity density corresponding to the uncertainty on the distance in GW170608 (i.e. $L_{\rm 1.4\,GHz}\approx (2.1-12)\times10^{28}$\,erg\,s$^{-1}$\,Hz$^{-1}$). 
As evident from this Figure, for a jet of energy $E_{\rm jet}=10^{49}$~ergs, only jet angles $\theta_{\rm jet}\gtrsim 40^\circ$
are compatible with the full range for the flux limit. It is interesting to discuss this limit in light of the constraints
derived	by \citet{Perna2018} for GW150914. The fact that GW150914
was the	only $\gamma$-ray candidate out of 10 BBH detections in the
O1/O2 observing runs led to the limit $(E_{\rm iso}/10^{49}{\rm
erg})(\theta_{\rm jet}/20^\circ)\lesssim 1$.
In the case of GW170608, the constraint	$\theta_{\rm jet}\gtrsim 40^\circ$
with $E_{\rm jet}=10^{49}$\,ergs is equivalent to $E_{\rm iso} \lesssim
4.2\times 10^{49}$~erg,	which is still compatible with the statistical
constraint previously derived from GW150914.

We should finally note that the afterglow models here presented are computed for a typical interstellar density of $n=0.01$~cm$^{-3}$ \citep{Perna2018}.
However, the flux brightness scales roughly as $n^{1/2}$ for a range of conditions \citep{Sari1998}. Hence, merger events
in lower densities could be more energetic and still be below the observational limits, while mergers in denser regions
would be constrained to being less  energetic.

\section{Summary and conclusion}
\label{sec:conclusion}
We have presented radio follow-up observations of a \textit{Fermi}/LAT candidate transient identified in the error region of GW170608, a BBH merger discovered by the LIGO detectors. Our observations with the VLA at 1.4\,GHz were complemented with VLITE data at 338\,MHz. We identify 25 sources with $SNR\gtrsim 10$ in the crowded VLA images of the LAT field. A few of these also had a counterpart with $SNR\gtrsim 5$ in VLITE data. Over the three epochs of our VLA follow up, none of the sources showed evidence for significant variability. Based on the lack of variability, we conclude that it is very unlikely that any of the VLA sources in our field are associated with the radio afterglow of the \textit{Fermi}/LAT candidate burst. 

We have compared the limits derived from our VLA observations with theoretical expectations for the radio afterglows potentially associated with jets launched in BBH mergers \citep{Perna2019}. Altogether, our analysis shows the key role that broad-band follow ups to GW events can play to gradually restrict the allowed parameter space  for electromagnetically bright, relativistic outflows driven by the merger of two BHs.

With this study we have demonstrated the feasibility of radio follow-up observations that, in the near-future observing runs of the advanced LIGO and Virgo detectors, could clarify the fundamental physics question of whether BBH mergers can be accompanied  by  relativistic outflows powering GRB-like bursts and afterglows. With the number of well-localized BBH mergers destined to increase thanks to the improving sensitivity of ground-based GW detectors, and with the Virgo interferometer joining the two LIGO detectors for longer observing runs \citep{LivRev}, it is crucial that the community  maintains active follow-up efforts of nearby BBHs. If appropriately planned, these follow-up efforts may ultimately help us identify a BBH afterglow, or set constraining upper-limits on the numerous models that have been proposed in the literature to explain the still largely debated, possible association between GW150914 and a \textit{Fermi}/GBM $\gamma$-ray transient. 

We finally note that commensal VLITE observation could also uncover potential coherent radio emission generated near the instant of merger. The last, in turn, could probe the immediate magnetic environment and the properties of the intergalactic medium \citep{Callister2019}.

\acknowledgments K.A., R.S., and A.C. acknowledge support from the NSF CAREER award \#1455090. R.P. acknowledges support from the NSF under grant AST-1616157. Basic research in radio astronomy at the Naval Research Laboratory is supported by 6.1 Base funding. The Center for Computational Astrophysics at the
Flatiron Institute is supported by the Simons Foundation. The National Radio Astronomy Observatory is a facility of the National Science Foundation operated under cooperative agreement by Associated Universities, Inc.  We thank Dale Frail, Mansi Kasliwal, Namir Kassim, and Nipuni Palliyaguru for constructive comments. 

\bibliography{biblio}

\begin{table*}[h]
\begin{center}\begin{footnotesize}
\caption{Results of the three epochs of our VLA follow-up campaign of GW170608 at 1.4 GHz. Columns are, from left to right: source name; R.A. and Dec. from our VLA images; object class (RadioS for all sources with a counterpart within 2\,\arcmin of the VLA position in FIRST (F) or NVSS (N), or object class of the closest source within 2\,\arcmin found in NED if no NVSS/FIRST counterpart is present); Modified Julian Date (MJD) of our VLA observations; epoch in days since GW170608 discovery; VLA flux density; NVSS or FIRST flux density (when available); offset between the source location as measured in our VLA images and that reported in NVSS or FIRST,  or (if not available) offset of the closest source reported in NED; VLA position error; NVSS or FIRST position error. \label{tb:VLASources}}
\begin{tabular}{ccccccccccc}
\hline
\hline
Name &  R.A.\,Dec.  &  Class& Epoch & $\Delta T$  & $F_{\rm 1.4\,GHz}$ &$F_{\rm 1.4\,GHz}$& Offset & Pos.Err. & Pos.Err.\\
&  (VLA)  &  &  &  & (VLA) &  (NVSS/ &  &   (VLA)& (NVSS/\\
 &  &  &  &  &  &  FIRST) & &  &   FIRST)\\
& (hh:mm:ss\,deg:mm:ss) & & (MJD) & (day) & (mJy)& (mJy) & (\arcsec) & ( \arcsec)& ( \arcsec)\\
\hline
S1   &08:33:16.15\,+43:23:51.01& RadioS(F) & 57917.333 & 4 &3.12 $\pm$ 0.40&3.66 $\pm$ 0.22& 0.30&0.35 & 0.11\\
     &"~~~~~~~~~~~~~~~~~"      &   "    & 57933.500 & 21  &3.16 $\pm$ 0.17& " &" &0.44 & "\\
     &"~~~~~~~~~~~~~~~~~"      &   "    & 57983.500 & 67  &2.86 $\pm$ 0.16& " &" &0.47 & "\\
\hline
S2   &08:31:26.13\,+43:25:12.49& RadioS(N) & 57917.333 & 4 &98.4 $\pm$ 4.9&149 $\pm$ 4.5& 0.52&0.34 & 1.50\\
     &"~~~~~~~~~~~~~~~~~"      &    "   & 57933.500 & 21  &110.0 $\pm$ 5.5& " &"      &0.41 & "\\
     &"~~~~~~~~~~~~~~~~~"      &    "   & 57983.500 & 67  &109.1 $\pm$ 5.5& " &"      &0.43 & "\\
\hline
S3   &08:32:43.30\,+43:14:36.05& RadioS(N) & 57917.333 & 4 &31.5 $\pm$ 1.6&43.0 $\pm$ 1.4& 0.94&0.34 & 0.5\\
     &"~~~~~~~~~~~~~~~~~"      &    "   & 57933.500 & 21  &33.4 $\pm$ 1.7& " &"      &0.41 & "\\
     &"~~~~~~~~~~~~~~~~~"      &    "   & 57983.500 & 67  &31.7 $\pm$ 1.6& " &"      &0.43 & "\\
\hline
S4   &08:32:29.39\,+43:02:58.04& RadioS(N) & 57917.333 & 4 &24.7 $\pm$ 1.3&196.6 $\pm$ 4.6& 6.3&0.35 & 1.5\\
     &"~~~~~~~~~~~~~~~~~"      &    "   & 57933.500 & 21  &26.6 $\pm$ 1.4& " &"      &0.42 & "\\
     &"~~~~~~~~~~~~~~~~~"      &    "   & 57983.500 & 67  &29.5 $\pm$ 1.5& " &"      &0.44 & "\\
\hline
S5   &08:31:11.34\,+43:16:06.66& RadioS(F) & 57917.333 & 4 &11.6 $\pm$ 0.1&10.4 $\pm$ 0.54& 0.60&0.72 & 0.1\\
     &"~~~~~~~~~~~~~~~~~"      &    "   & 57933.500 & 21  &13.4 $\pm$ 0.7& " &"      &0.42 & "\\
     &"~~~~~~~~~~~~~~~~~"      &    "   & 57983.500 & 67  &12.8 $\pm$ 0.7& " &"      &0.44 & "\\
\hline
S6   &08:32:28.84\,+43:03:20.21& RadioS(F) & 57917.333 & 4 &17.99 $\pm$ 0.93&22.5 $\pm$ 1.1& 1.5&0.35 & 0.1\\
     &"~~~~~~~~~~~~~~~~~"      &    "   & 57933.500 & 21  &22.83 $\pm$ 1.16& " &"      &0.42 & "\\
     &"~~~~~~~~~~~~~~~~~"      &    "   & 57983.500 & 67  &18.29 $\pm$ 0.95& " &"      &0.45 & "\\
\hline
S7   &08:31:00.04\,+43:27:28.96& RadioS(F) & 57917.333 & 4 &9.22 $\pm$ 0.16&12.61 $\pm$ 0.65& 1.3&0.53 & 0.1\\
     &"~~~~~~~~~~~~~~~~~"      &    "   & 57933.500 & 21&8.96 $\pm$ 0.45& " &" &0.42 & "\\
     &"~~~~~~~~~~~~~~~~~"      &    "   & 57983.500 & 67&8.82 $\pm$ 0.46& " &" &0.45 & "\\
\hline
S8   &08:31:14.21\,+43:11:19.64& RadioS(F) & 57917.333 & 4 &2.11 $\pm$ 0.16&2.45 $\pm$ 0.18&  0.7&0.95 & 0.16\\
     &"~~~~~~~~~~~~~~~~~"      &    "   & 57933.500 & 21  &2.76 $\pm$ 0.17& " &"      &0.70 & "\\
     &"~~~~~~~~~~~~~~~~~"      &    "   & 57983.500 & 67  &2.32 $\pm$ 0.18& " &"      &0.79 & "\\
\hline
S9   &08:32:56.03\,+43:13:24.65& RadioS(F) & 57917.333 & 4 &5.43 $\pm$ 0.47&6.19 $\pm$ 0.34& 0.6&0.35 & 0.1\\
     &"~~~~~~~~~~~~~~~~~"      &    "   & 57933.500 & 21  &4.60 $\pm$ 0.24& " &"      &0.43 & "\\
     &"~~~~~~~~~~~~~~~~~"      &    "   & 57983.500 & 67  &4.95 $\pm$ 0.26& " &"      &0.45 & "\\
\hline
S10   &08:33:50.53\,+43:20:39.09& RadioS(F) & 57917.333 & 4 &7.82 $\pm$ 0.29&3.55 $\pm$ 0.23& 1.3&0.37 & 0.29\\
     &"~~~~~~~~~~~~~~~~~"      &    "   & 57933.500 & 21  &7.52 $\pm$ 0.39& " &"      &0.43 & "\\
     &"~~~~~~~~~~~~~~~~~"      &    "   & 57983.500 & 67  &7.16 $\pm$ 0.37& " &"      &0.45 & "\\
\hline
S11   &08:33:48.59\,+43:20:04.31& RadioS(F) & 57917.333 & 4 &5.14 $\pm$ 0.19&2.15 $\pm$ 0.18& 1.0&0.42 & 0.57\\
     &"~~~~~~~~~~~~~~~~~"      &    "   & 57933.500 & 21  &4.50 $\pm$ 0.25& " &"      &0.46 & "\\
     &"~~~~~~~~~~~~~~~~~"      &    "   & 57983.500 & 67  &4.86 $\pm$ 0.27& " &"      &0.48 & "\\
\hline
S12  &08:31:20.18\,+43:33:00.13& RadioS(F) & 57917.333 & 4 &3.62 $\pm$ 0.28&3.9 $\pm$ 0.2& 1.4&0.37 & 0.11\\
     &"~~~~~~~~~~~~~~~~~"      &    "   & 57933.500 & 21  &3.61 $\pm$ 0.20& " &"      &0.47 & "\\
     &"~~~~~~~~~~~~~~~~~"      &    "   & 57983.500 & 67  &4.13 $\pm$ 0.23& " &"      &0.49 & "\\
\hline
S13  &08:33:05.09\,+43:13:37.07& RadioS(F) & 57917.333 & 4 &2.14 $\pm$ 0.13&2.44 $\pm$ 0.18& 0.60&0.42 & 0.19\\
     &"~~~~~~~~~~~~~~~~~"      &    "   & 57933.500 & 21  &2.26 $\pm$ 0.13& " &"      &0.48 & "\\
     &"~~~~~~~~~~~~~~~~~"      &    "   & 57983.500 & 67  &1.99 $\pm$ 0.13& " &"      &0.55 & "\\
\hline
S14  &08:32:45.34\,+43:29:49.34& RadioS(F) & 57917.333 & 4 &1.59 $\pm$ 0.19&1.81 $\pm$ 0.16& 0.40&0.36 & 0.20\\
     &"~~~~~~~~~~~~~~~~~"      &    "   & 57933.500 & 21  &1.48 $\pm$ 0.09& " &"      &0.50 & "\\
     &"~~~~~~~~~~~~~~~~~"      &    "   & 57983.500 & 67  &1.31 $\pm$ 0.09& " &"      &0.57 & "\\
\hline
S15  &08:32:08.63\,+43:40:05.18& RadioS(N) & 57917.333 & 4 &3.65 $\pm$ 0.15&3.8 $\pm$ 0.5& 3.1&0.66 & 9.1\\
     &"~~~~~~~~~~~~~~~~~"      &    "   & 57933.500 & 21&3.30 $\pm$ 0.21& " &"      &0.52 & "\\
     &"~~~~~~~~~~~~~~~~~"      &    "   & 57983.500 & 67&3.20 $\pm$ 0.21& " &"      &0.58 & "\\
\hline
S16  &08:32:59.90\,+43:40:03.57& RadioS(N) & 57917.333 & 4 &3.90 $\pm$ 0.23&4.00 $\pm$ 0.5& 0.88&0.44 & 11\\
     &"~~~~~~~~~~~~~~~~~"      &    "   & 57933.500 & 21&3.05 $\pm$ 0.20& " &"      &0.55 & "\\
     &"~~~~~~~~~~~~~~~~~"      &    "   & 57983.500 & 67&3.51 $\pm$ 0.23& " &"      &0.57 & "\\
\hline
S17  &08:31:53.81\,+43:27:20.12& Galaxy & 57917.333 & 4 &0.99 $\pm$ 0.20& - & 1.6&0.36 & -\\
     &"~~~~~~~~~~~~~~~~~"      &    "   & 57933.500 & 21&1.05 $\pm$ 0.07& - &"      &0.57 & -\\
     &"~~~~~~~~~~~~~~~~~"      &    "   & 57983.500 & 67&0.65 $\pm$ 0.07& - &"      &0.86 & -\\
\hline
S18  &08:32:24.99\,+43:18:26.14& Galaxy & 57917.333 & 4 &0.67 $\pm$ 0.06& - & 0.72&0.52 & -\\
     &"~~~~~~~~~~~~~~~~~"      &    "   & 57933.500 & 21&0.88 $\pm$ 0.06& - &"      &0.59 & -\\
     &"~~~~~~~~~~~~~~~~~"      &    "   & 57983.500 & 67&0.64 $\pm$ 0.06& - &"      &0.83 & -\\
\hline
S19  &08:33:04.28\,+43:32:37.78& UvS & 57917.333 & 4 &1.12 $\pm$ 0.09& -& 7.7 &0.60 & -\\
     &"~~~~~~~~~~~~~~~~~"      &    "   & 57933.500 & 21&0.98 $\pm$ 0.08& - &"      &0.70 & -\\
     &"~~~~~~~~~~~~~~~~~"      &    "   & 57983.500 & 67&0.99 $\pm$ 0.09& - &"      &0.78 & -\\
\hline
S20  &08:32:02.84\,+43:29:45.43& Galaxy & 57917.333 & 4 &0.89 $\pm$ 0.07& - & 1.1  &0.55 & -\\
     &"~~~~~~~~~~~~~~~~~"      &    "   & 57933.500 & 21&0.75 $\pm$ 0.06& - &"      &0.70 & -\\
     &"~~~~~~~~~~~~~~~~~"      &    "   & 57983.500 & 67&0.85 $\pm$ 0.07& - &"      &0.74 & -\\
\hline
\end{tabular}
\end{footnotesize}\end{center}\end{table*}
     
\begin{table*}[h]
\begin{center}\begin{footnotesize}
\begin{tabular}{ccccccccccc}
\hline
Name &  R.A.\,Dec.  &  Class& Epoch & $\Delta T$  & $F_{\rm 1.4\,GHz}$ &$F_{\rm 1.4\,GHz}$& Offset & Pos.Err. & Pos.Err.\\
&  (VLA)  &  &  &  &  (VLA)& (NVSS/ &  &   (VLA)& (NVSS/\\
 &  &  &  &  &  &   FIRST)& &  &   FIRST)\\
& (hh:mm:ss\,deg:mm:ss) & & (MJD) & (day) & (mJy)& (mJy) & (\arcsec) & ( \arcsec)& ( \arcsec)\\
\hline
S21  &08:32:12.72\,+43:23:06.47& RadioS(F) & 57917.333 & 4 &1.61 $\pm$ 0.10& 1.29 $\pm$ 0.15 & 0.70&0.39 & 0.65\\
     &"~~~~~~~~~~~~~~~~~"      &   & 57933.500 & 21  &1.72 $\pm$ 0.10& "& "&0.46 & "\\
     &"~~~~~~~~~~~~~~~~~"      &   & 57983.500 & 67  &1.70 $\pm$ 0.10& "& "&0.50 & "\\
\hline
S22  &08:32:05.46\,+43:17:27.99& RadioS(F) & 57917.333 & 4 &1.69 $\pm$ 0.09& 2.01 $\pm$ 0.17 & 0.50&0.42 & 0.26\\
     &"~~~~~~~~~~~~~~~~~"      &   & 57933.500 & 21 &1.84 $\pm$ 0.10& "& "&0.47 & "\\
     &"~~~~~~~~~~~~~~~~~"      &   & 57983.500 & 67  &1.76 $\pm$ 0.10& "& "&0.51 & "\\
\hline
S23  &08:32:19.72\,+43:16:11.42& RadioS(F) & 57917.333 & 4 &1.49 $\pm$ 0.07& 1.73 $\pm$ 0.16 & 0.60&0.56 & 0.28\\
     &"~~~~~~~~~~~~~~~~~"      &   & 57933.500 & 21  &1.54 $\pm$ 0.09& "& "&0.49 & "\\
     &"~~~~~~~~~~~~~~~~~"      &   & 57983.500 & 67  &1.44 $\pm$ 0.09& "& "&0.55 & "\\
\hline
S24  &08:32:43.90\,+43:10:43.91&  Galaxy & 57917.333 & 4 &0.85 $\pm$ 0.11& -& 14&0.49 & -\\
     &"~~~~~~~~~~~~~~~~~"      &   & 57933.500 & 21  &0.92 $\pm$ 0.09& -& "&0.79 & -\\
     &"~~~~~~~~~~~~~~~~~"      &   & 57983.500 & 67  &0.88 $\pm$ 0.10& -& "&0.94 & -\\
\hline
S25   &08:32:35.05\,+43:33:46.81& Star  & 57917.333 & 4 &0.87 $\pm$ 0.08& -& 7.4&0.62 & -\\
     &"~~~~~~~~~~~~~~~~~"      &        & 57933.500 & 21  &0.67 $\pm$ 0.07& -& "&0.87 & -\\
     &"~~~~~~~~~~~~~~~~~"      &        & 57983.500 & 67  &0.78 $\pm$ 0.08& -& "&0.89 & -\\
\hline
\end{tabular}
\end{footnotesize}\end{center}\end{table*}

\end{document}